\documentclass[twoside,slac_one]{revtex4}
\pdfoutput=1

\usepackage{xspace}
\usepackage{graphicx}
\usepackage{fancyhdr}
\usepackage{amsmath} 
\usepackage{bm}
\usepackage{amsxtra}
\usepackage{amssymb}
\usepackage{amsthm}
\usepackage{latexsym}
\usepackage{lscape}

\def\lhcb {LHC{\em b\/}\xspace}
\def\btev {BTe\kern -0.1em V}

\def\babar{\mbox{\slshape B\kern-0.1em{\small A}\kern-0.1em B\kern-0.1em{\small A\kern-0.2em R}}\xspace}

\def\mup        {\ensuremath{\mu^+}\xspace}
\def\mun        {\ensuremath{\mu^-}\xspace} 
\def\mumu       {\ensuremath{\mu^+\mu^-}\xspace}

\def\ellell     {\ensuremath{\ell^+ \ell^-}\xspace}

\def\g     {\ensuremath{\gamma}\xspace}

\def\Z      {\ensuremath{Z^0}\xspace}

\def\ccbar {\ensuremath{c\overline c}\xspace}

\def\pim   {\ensuremath{\pi^-}\xspace}

\def\pimp  {\ensuremath{\pi^\mp}\xspace}

\def\Kbar  {\kern 0.2em\overline{\kern -0.2em K}{}\xspace}

\def\Kz    {\ensuremath{K^0}\xspace}
\def\Kzb   {\ensuremath{\Kbar^0}\xspace}
\def\KzKzb {\ensuremath{\Kz \kern -0.16em \Kzb}\xspace}
\def\Kp    {\ensuremath{K^+}\xspace}
\def\Km    {\ensuremath{K^-}\xspace}
\def\Kpm   {\ensuremath{K^\pm}\xspace}
\def\Kmp   {\ensuremath{K^\mp}\xspace}
\def\KpKm  {\ensuremath{\Kp \kern -0.16em \Km}\xspace}

\def\Kstarz  {\ensuremath{K^{*0}}\xspace}
\def\Kstarzb {\ensuremath{\Kbar^{*0}}\xspace}

\def\Dbar    {\kern 0.2em\overline{\kern -0.2em D}{}\xspace}

\def\Dz      {\ensuremath{D^0}\xspace}
\def\Dzb     {\ensuremath{\Dbar^0}\xspace}
\def\DzDzb   {\ensuremath{\Dz {\kern -0.16em \Dzb}}\xspace}
\def\Dp      {\ensuremath{D^+}\xspace}
\def\Dm      {\ensuremath{D^-}\xspace}

\def\DpDm    {\ensuremath{\Dp {\kern -0.16em \Dm}}\xspace}

\def\B       {\ensuremath{B}\xspace}
\def\Bbar    {\kern 0.18em\overline{\kern -0.18em B}{}\xspace}

\def\Bz      {\ensuremath{B^0}\xspace}
\def\Bzb     {\ensuremath{\Bbar^0}\xspace}
\def\BzBzb   {\ensuremath{\Bz {\kern -0.16em \Bzb}}\xspace}
\def\Bu      {\ensuremath{B^+}\xspace}
\def\Bub     {\ensuremath{B^-}\xspace}

\def\BpBm    {\ensuremath{\Bu {\kern -0.16em \Bub}}\xspace}
\def\Bd      {\ensuremath{B_d}\xspace}
\def\Bs      {\ensuremath{B_s}\xspace}

\def\Bdb     {\ensuremath{\Bbar_d}\xspace}

\def\jpsi     {\ensuremath{{J\mskip -3mu/\mskip -2mu\psi\mskip 2mu}}\xspace}

\mathchardef\Upsilon="7107
\def\Y#1S{\ensuremath{\Upsilon{(#1S)}}\xspace}

\mathchardef\Deltares="7101
\mathchardef\Xi="7104
\mathchardef\Lambda="7103
\mathchardef\Sigma="7106
\mathchardef\Omega="710A

\def\Deltabar{\kern 0.25em\overline{\kern -0.25em \Deltares}{}\xspace}
\def\Lbar{\kern 0.2em\overline{\kern -0.2em\Lambda\kern 0.05em}\kern-0.05em{}\xspace}
\def\Sigbar{\kern 0.2em\overline{\kern -0.2em \Sigma}{}\xspace}
\def\Xibar{\kern 0.2em\overline{\kern -0.2em \Xi}{}\xspace}
\def\Obar{\kern 0.2em\overline{\kern -0.2em \Omega}{}\xspace}
\def\Nbar{\kern 0.2em\overline{\kern -0.2em N}{}\xspace}
\def\Xb{\kern 0.2em\overline{\kern -0.2em X}{}\xspace}

\def\btosgam    {\ensuremath{b \to s \g}\xspace}

\def\btosll     {\ensuremath{b \to s \ellell}\xspace}

\def\pt         {\mbox{$p_T$}\xspace}

\newcommand{\tev}{\ensuremath{\mathrm{\,Te\kern -0.1em V}}\xspace}
\newcommand{\gev}{\ensuremath{\mathrm{\,Ge\kern -0.1em V}}\xspace}
\newcommand{\mev}{\ensuremath{\mathrm{\,Me\kern -0.1em V}}\xspace}
\newcommand{\kev}{\ensuremath{\mathrm{\,ke\kern -0.1em V}}\xspace}
\newcommand{\ev}{\ensuremath{\mathrm{\,e\kern -0.1em V}}\xspace}
\newcommand{\gevc}{\ensuremath{{\mathrm{\,Ge\kern -0.1em V\!/}c}}\xspace}
\newcommand{\mevc}{\ensuremath{{\mathrm{\,Me\kern -0.1em V\!/}c}}\xspace}
\newcommand{\gevcc}{\ensuremath{{\mathrm{\,Ge\kern -0.1em V\!/}c^2}}\xspace}
\newcommand{\gevgevcccc}{\ensuremath{{\mathrm{\,Ge\kern -0.1em V^2\!/}c^4}}\xspace}
\newcommand{\mevcc}{\ensuremath{{\mathrm{\,Me\kern -0.1em V\!/}c^2}}\xspace}

\def\mm   {\ensuremath{\rm \,mm}\xspace}

\def\mum  {\ensuremath{\,\mu\rm m}\xspace}

\def\invpb {\ensuremath{\mbox{\,pb}^{-1}}\xspace}

\def\invfb   {\ensuremath{\mbox{\,fb}^{-1}}\xspace}

\def\mus  {\ensuremath{\rm \,\mus}\xspace}

\def\mus        {\ensuremath{\,\mu{\rm s}}\xspace}    

\def\cmsqs         {\ensuremath{{\rm \,cm}^{-2} {\rm s}^{-1}}\xspace}

\def\to                 {\ensuremath{\rightarrow}\xspace}

\def\pep2{PEP-II}

\def\gsim{{~\raise.15em\hbox{$>$}\kern-.85em
          \lower.35em\hbox{$\sim$}~}\xspace}
\def\lsim{{~\raise.15em\hbox{$<$}\kern-.85em
          \lower.35em\hbox{$\sim$}~}\xspace}

\def\BdKsmm{\ensuremath{\Bd \to \Kstarz\mup\mun}\xspace}

\def\BdKsg	{\ensuremath{\Bd \to \Kstarz \g}\xspace}

\begin{document}

\title{Electroweak penguin decays at \lhcb}

%

\author{T.~Blake on behalf of the \lhcb collaboration}
\affiliation{CERN, Switzerland}

\begin{abstract}
Promising ways to search for New Physics effects in radiative penguin decays are in the angular analysis of \BdKsmm, in the measurement of direct CP violation in \BdKsg  and a time dependent analysis of $\Bs\to \phi \g$. All of these studies are being pursued at LHCb. First results will be shown from the 2010 and early 2011 data, with particular emphasis on \BdKsmm. 
\end{abstract}

\maketitle

\thispagestyle{fancy}


\section{Introduction}

The decay processes \btosll and \btosgam are flavour changing neutral currents that are forbidden at tree level in the Standard Model (SM). These processes can proceed via higher order electroweak \Z/\g penguin or box diagrams. In extensions to the SM, new virtual particles can enter in competing (loop order) diagrams, leading, in the absence of a dominant SM tree process, to comparably large deviations from SM predictions. These deviations may be: in enhancements (or supprssion) of branching fractions; in angular distributions (e.g. \BdKsmm); CP or Isospin aymmetries. Rare decay processes provide a complementary approach to direct searches at the general purpose detectors and can provide sensitivity to new particles with masses of up-to $\mathcal{O}(10-100\tev)$. 

The angular analysis of \BdKsmm, based on 309\invpb of integrated luminosity collected by the \lhcb experiment in 2011, is described in Sec.~\ref{sec:btosll}. More details on the angular analysis of \BdKsmm with this data set can be found in Ref.~\cite{bib:kstmm}. Initial studies of \BdKsg and $\Bs\to\phi\g$ based on  88\invpb of integrated collected in 2010 and 2011 are described in Sec.~\ref{sec:btosgam}. The search for the rare decay $\Bs\to\mumu$ is described in detail elsewhere in these proceedings~\cite{bib:bstomumu}.

\subsection{The \lhcb detector}

The \lhcb detector is a single-arm spectrometer designed to study $b$-hadron decays with an acceptance for charged tracks with pseudorapidity, $\eta$, of $2<\eta<5$. Primary proton-proton vertices~(PVs), and secondary \B-vertices are identified in a silicon strip vertex detector (the VELO) that approaches within 8\mm of the LHC beam. Tracks from charged particles are reconstructed in the vertex detector and a set of tracking stations and their curvature in the dipole magnet allows momenta to be determined with a precision of $\delta p/p =0.35$--$0.5\%$. Two Ring Imaging CHerenkov (RICH) detectors allow for charged hadrons to be separated over a momentum range $2\gevc<p<100\gevc$. Muons wth momentum above 3\gevc are identified on the basis of the number of hits left in detectors interleaved with an iron muon filter. Further details on the \lhcb detector can be found in Ref.~\cite{DetectorPaper}. Key for the rare electroweak penguin processes are the excellent momentum and mass resolution provided by the long lever arm for tracking, the primary-secondary vertex separation provided by the VELO and the ability to reject a range of exclusive backgrounds provided by \lhcb's RICH detectors.

\subsection{\lhcb data taking performance in 2011} 

In the first three months of data taking in 2011, \lhcb accumulated $\sim300\invpb$ of integrated luminosity at $\sqrt{s} = 7\tev$. This luminosity was delivered by the LHC at instantaneous luminosities of $3\times 10^{32}\cmsqs$, 50\% above the original design luminosity of \lhcb. At this instantaneous luminosity, it is expected that \lhcb will collect $\mathcal{O}(1\invfb)$ of integrated luminosity in 2011. 

\section{The decay \BdKsmm}
\label{sec:btosll}

\begin{figure}[h]
\centering
\includegraphics[width=80mm]{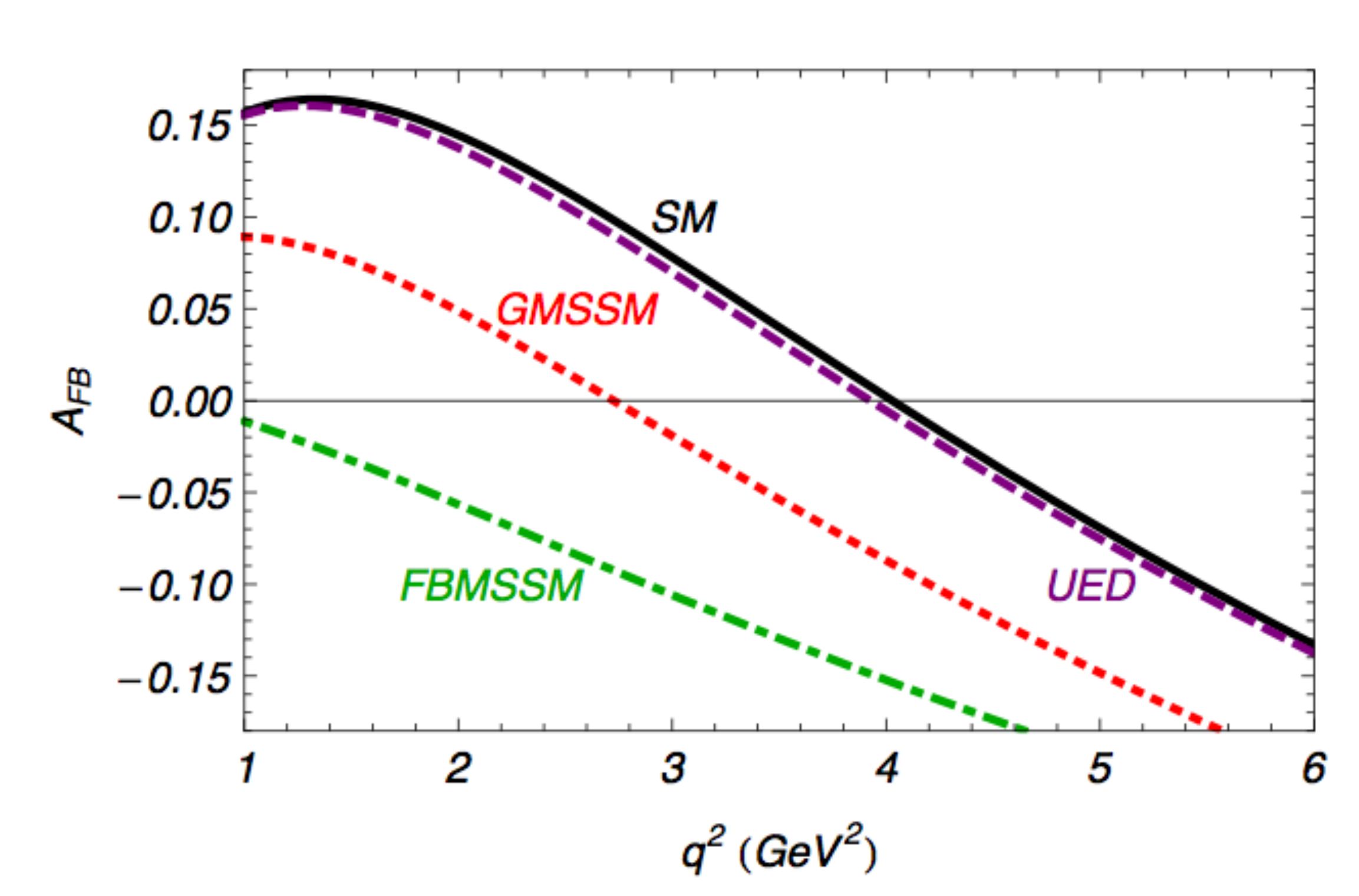} 
\caption{The forward backward asymmetry of muons in \BdKsmm  as a function of   $q^{2} = m_{\mumu}^{2}$ predicted in the SM  (solid-black line) and  a variety of NP models. The SM prediction and the NP models are described in detail in Ref.~\cite{bib:altmannshofer}. \label{fig:bdkstmm:predict} }
\end{figure}

The decay \BdKsmm is an example of a \btosll process that can be a highly sensitive probe of new right-handed currents and large contributions from new scalar or pseudo-scalar couplings. In many models these new virtual particles give rise to modifications in the distribution of the daughters of the \Bd that can be probed through an angular analysis. To achieve ultimate sensitivity, the long term goal of \lhcb is to perform a full angular analysis of the decay \cite{bib:kstarmumu:full}. However, with more modest data sets, the focus is instead on obseravables that are theoretically clean and  can be extracted from simple counting experiments or from simple fits to angular distributions. One such observable, that is widely discussed in the literature, is the forward-backward asymmetry of the muon system ($A_{FB}$). $A_{FB}$ varies with the invariant mass-squared of the dimuon pair ($q^{2}$) and in the SM changes sign at a well defined point, where the leading hadronic uncertainties cancel.  In many NP models the shape of $A_{FB}$ as a function of $q^{2}$ can be dramatically altered. The variation of $A_{FB}(q^{2})$ in the SM and several NP models in the low-$q^{2}$ region, from Ref.~\cite{bib:altmannshofer}, are shown in Fig.~\ref{fig:bdkstmm:predict}

Recent measurements from \babar~ \cite{bib:kstarmumu:babar}, Belle~\cite{bib:kstarmumu:belle} and CDF~\cite{bib:kstarmumu:cdf} have generated excitement as they appear to favour a forward-backward asymmetry with the opposite sign to the SM prediction at low-$q^{2}$ and no zero-crossing point.  

In 309\invpb \lhcb observes $302\pm20$ candidates in a $\pm50\mevcc$ $\Kpm\pimp\mumu$ mass window. Candidates were triggered by: a single high-$p_{T}$ muon in \lhcb's Level\,0 hardware trigger \cite{bib:level0}; a single high impact parameter and high-$p_{T}$ daughter \cite{bib:track} in the first stage of a two stage software trigger and `topologically' by partially reconstructing the \Bd decay in the second stage of the software trigger \cite{bib:topo}. 

Offline, candidates are first selected by applying a loose pre-selection based on the \Bd lifetime, daughter impact parameters and a requirement that the \Bd points back to one of the primary vertices in the event. A multivariate selected based on a Boosted Decision Tree (BDT) was then used to further reduce combinatorial background. The BDT combined information on the \Bd kinematics, \Bd vertex quality, and the kaon, pion and muon impact parameter and particle identification. The BDT allowed a signal-to-background ratio of three-to-one to be achieved across the $\pm50\mevcc$ signal mass window. The multivariate selection was trained using $\Bd\to\Kstarz\jpsi$ candidates and background from the upper mass sideband of a 36\invpb data set collected in 2010. These events are not used in the subsequent analysis. Care has also been taken to avoid further biasing the angular distribution in the offline selection.

Specific exclusive backgrounds from, e.g, $\Bd\to\jpsi (\to \mup \mun \{\pim\}) \Kstarz (\to \Kp \pim \{\mun \})$ where the pion from the \Kstarz is misidentified as a muon and a muon from the \jpsi as the pion, are removed by exchanging the mass assignments of the daughters and cutting on the the resulting \Bd mass in combination with cuts on the daughter particle identification. These selection criteria are almost 100\% efficient on genuine \BdKsmm candidates and reduce this type of background to a negligible level. In particular the exclusive background from \BdKsmm where the \Kstarz is misidentified as a \Kstarzb is reduced to 1\% of the level of the signal, diluting the measured $A_{FB}$ by 2\%. This dilution is accounted for in the results presented below. The regions around the \jpsi and $\psi(2S)$ are also removed as they are dominated by different physics from \ccbar loops (see Fig.~\ref{fig:bdkstmm:scatter}). 

The forward backward-asymmetry ($A_{FB}$) of the muon system is extracted from the angular distribution of the \mup (\mun) helicity angle w.r.t. the dimuon flight direction in the rest-frame of the \Bd (\Bdb). The probability density distribution for the cosine of this angle, $\cos\theta_{\ell}$ is given by: 

\begin{equation}
\frac{1}{\Gamma} \frac{\mathrm{d}^{2}\Gamma}{\mathrm{d}\cos\theta_{\ell}\,\mathrm{d}q^{2}} = \frac{3}{4}{F_{L}}(1-\cos^{2}\theta_{\ell}) + \frac{3}{8}(1-{F_{L}}) (1+\cos^{2}\theta_{\ell}) + {A_{FB}}\cos\theta_{\ell} 
\label{eq:costhetal}
\end{equation}

\begin{center}

\begin{figure}[!ht]
\centering
\includegraphics[width=130mm]{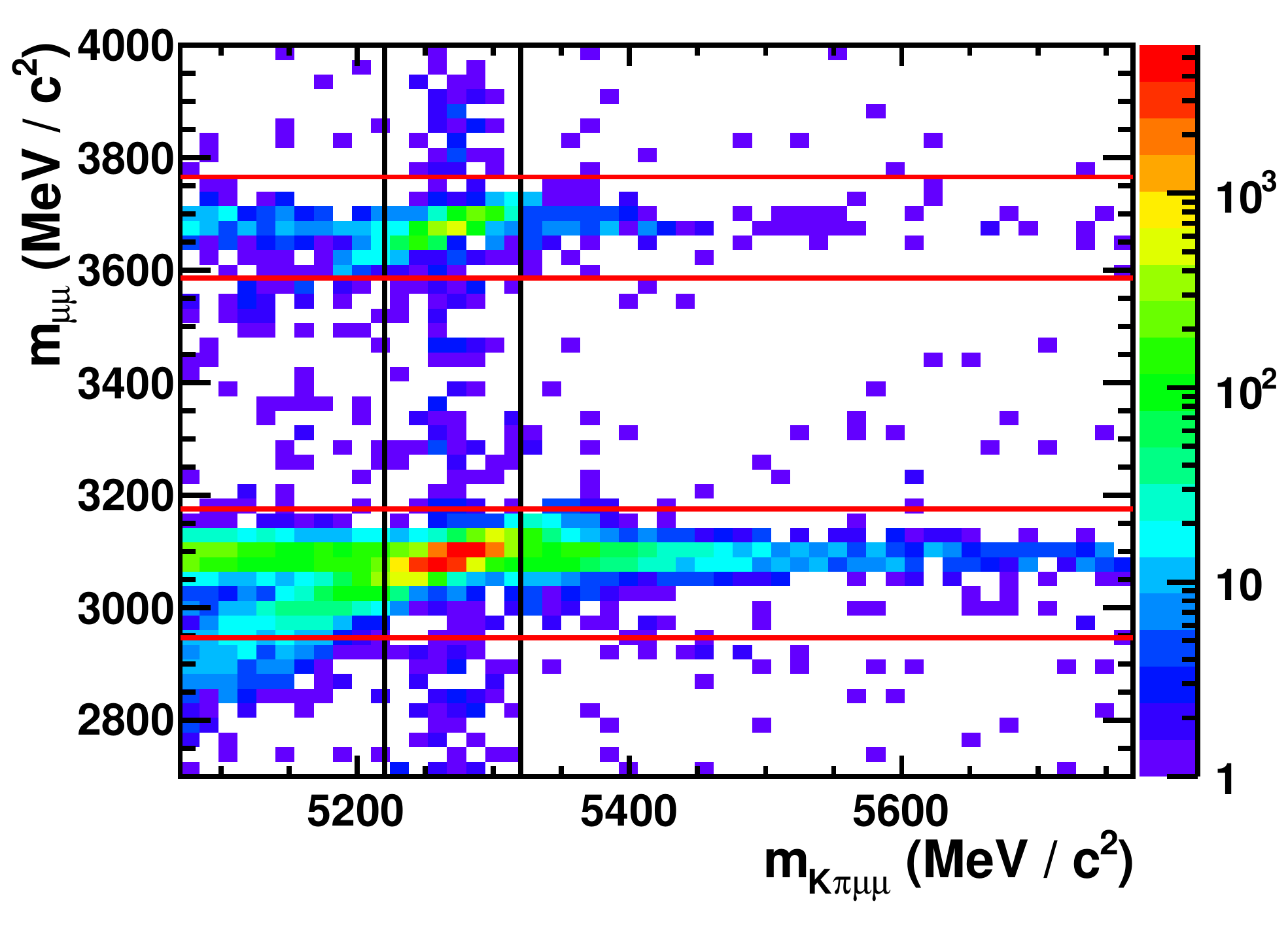} 
\caption{Scatter plot highlighting the correlation between $m_{\mumu}$ and $m_{K\pi\mum}$ for candidates used in the analysis. The vertical band illustrates the $\pm 50\mevcc$ signal mass window used in the analysis. The horizontal bands illustrate the \jpsi and $\psi(2S)$ regions taht are treated separately in the analysis. A clear \BdKsmm signal is visible across $m_{\mumu}$,  centered on the \Bd mass. \label{fig:bdkstmm:scatter}}
\end{figure}

\begin{figure}[!hb]
\centering
\includegraphics[width=130mm]{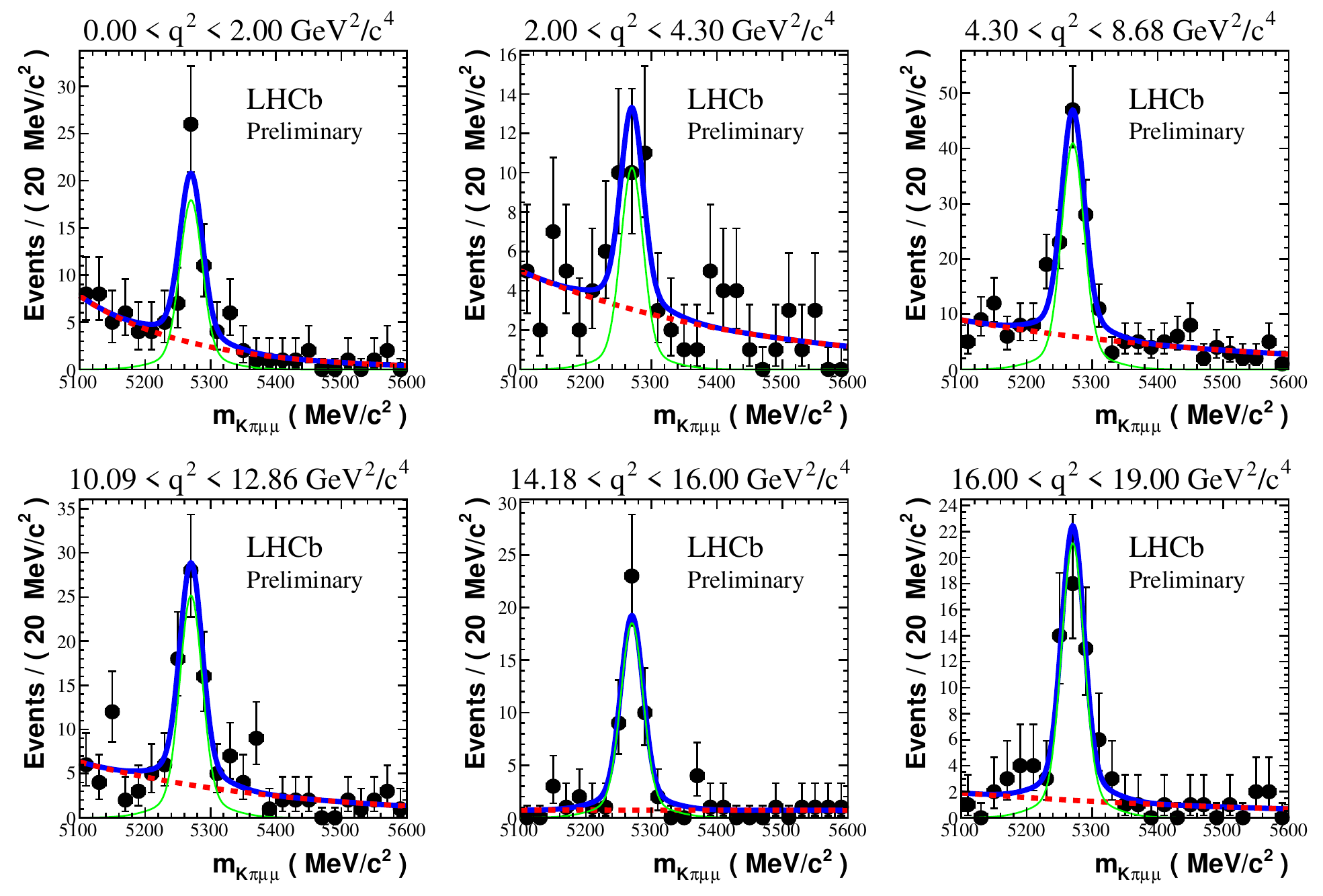}
\caption{The ${\Kpm\pimp\mu\mu}$ mass distribution of \BdKsmm candidates in six $q^{2}$ bins. The solid line shows a fit to this distribution with a double-Gaussian signal component (thin-green line) and Exponential background component (dashed-red line). } \label{fig:kstmm:mass}
\end{figure}

\end{center}

\noindent which also depends on $F_{L}$, the fraction of longitudinal polarisation of the \Kstarz. $F_{L}$ is constrained by simultaneously fitting the helicity angle of the kaon ($\theta_{K}$). The probability density function for $\cos\theta_{K}$ only depends on $F_{L}$ and is given by:

\begin{equation}
\frac{1}{\Gamma} \frac{\mathrm{d}^{2}\Gamma}{\mathrm{d}\cos\theta_{K}\,\mathrm{d}q^{2}} = \frac{3}{2}{F_{L}}\cos^{2}\theta_{K} + \frac{3}{4}(1-{F_{L}}) (1- \cos^{2}\theta_{K})
\label{eq:costhetak}
\end{equation}

The analysis is performed in six-$q^{2}$ bins, chosen for consistency to be the same binning scheme used by previous experiments. A significant \BdKsmm signal is visible in each of the six bins with an excellent signal-to-background ratio even in the lowest $q^{2}$ bin (see Fig.~\ref{fig:kstmm:mass}). 

Unfortunately the extraction of $A_{FB}$ is complicated by a correlation between $A_{FB}$ and $F_{L}$ that prevents the $A_{FB}$ from being large if $F_{L}$ is also large. The allowed region of phase space, where Eq.~\ref{eq:costhetal} remains positive and well defined, corresponds to $\left| A_{FB} \right| \leq \frac{3}{4}( 1- F_{L}$). To account for the `physical' region a profile likelihood scan is made over the plane, with a flat prior that the maximum likelihood must lie in the physical region. The statistical uncertainty on the central value of $A_{FB}$ and $F_{L}$ is calculated by integrating the likelihood to yield an (asymmetric) 68\% confidence limt on $A_{FB}$ and $F_{L}$.

\begin{figure}[h]
\centering
\includegraphics[width=80mm]{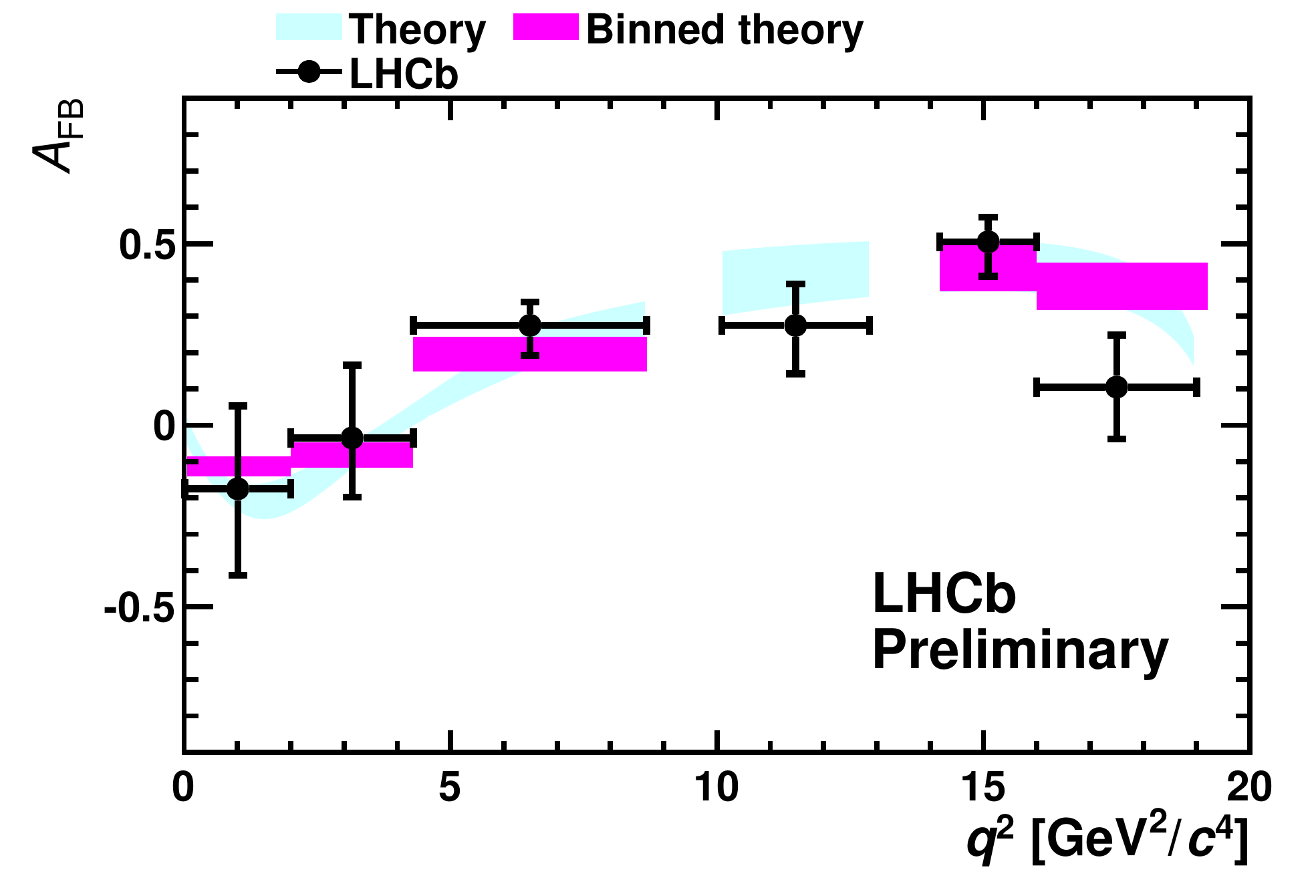} \\
\includegraphics[width=80mm]{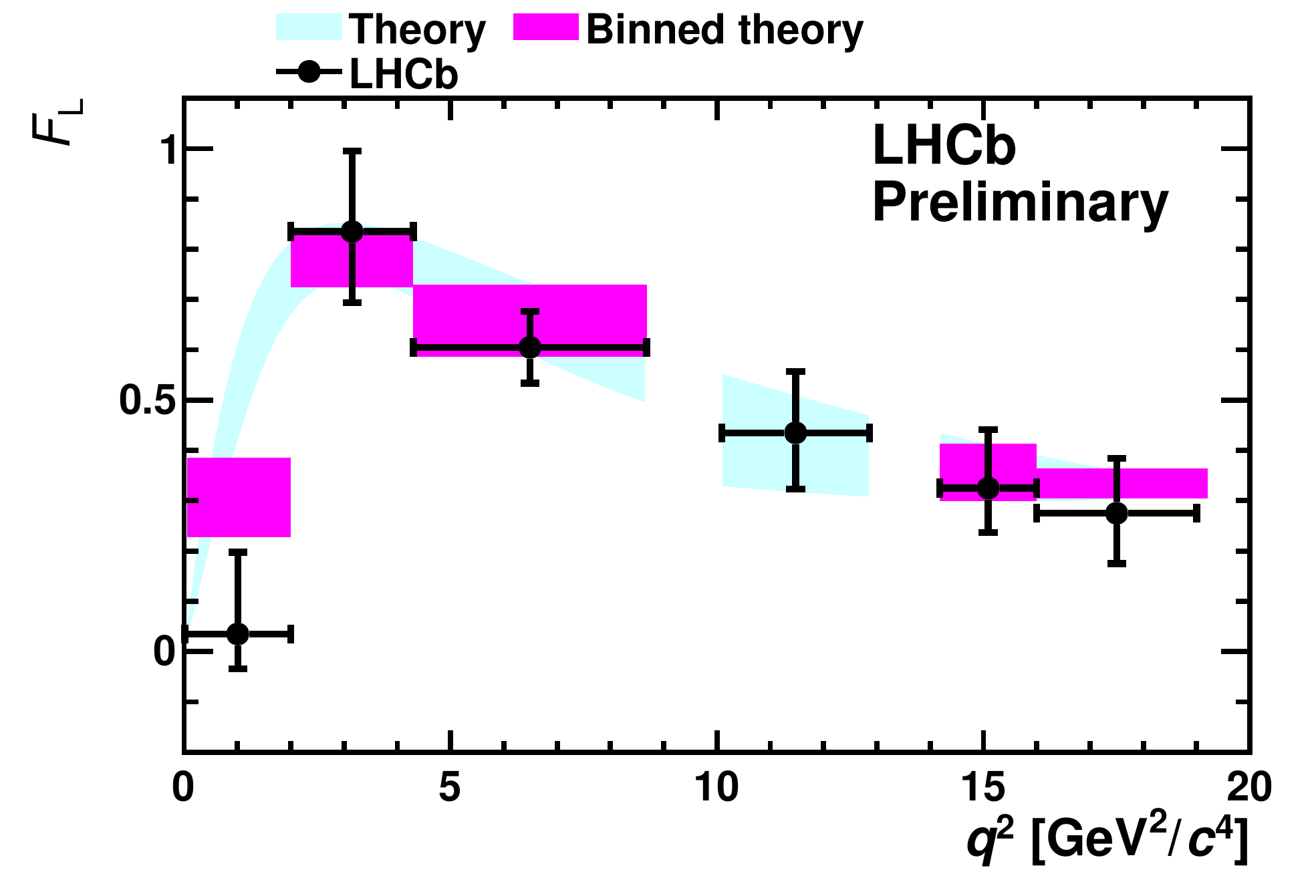} \\
\includegraphics[width=80mm]{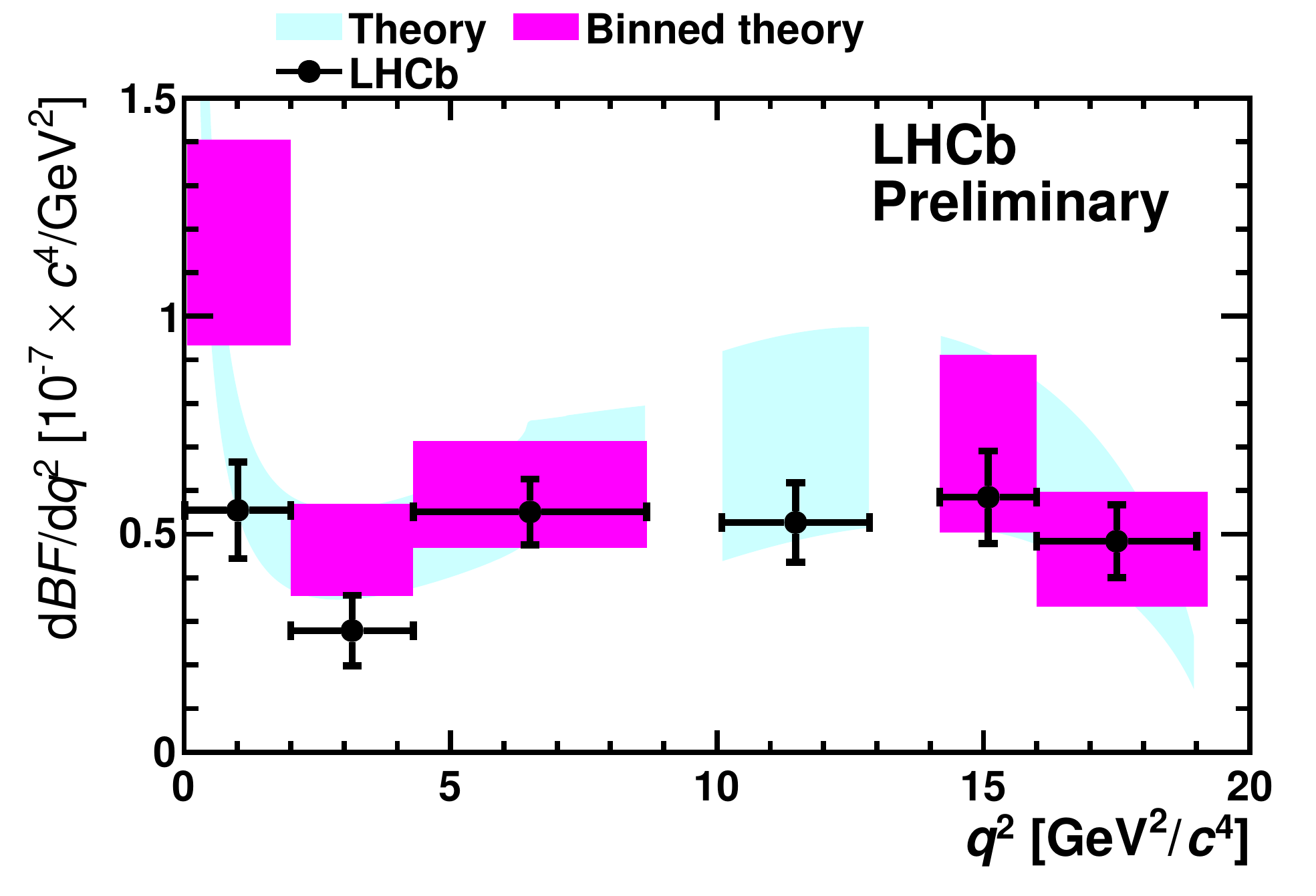} 
\caption{$A_{FB}$, $F_{L}$ and the differential branching fraction of \BdKsmm candidates as a function of $q^{2}$ in the six $q^{2}$ bins. The theory predictions are described in Ref.~\cite{Bobeth:2011gi}} \label{fig:bdkstmm:result}
\end{figure}

The fit results for $A_{FB}$ and $F_{L}$ in six $q^{2}$ bins are shown in Fig.~\ref{fig:bdkstmm:result} along with the differential branching fraction of \BdKsmm as a function of $q^{2}$. The differential branching fraction is extracted from fits to the $\Kpm\pimp\mumu$ mass distribution in the $q^{2}$ bins and normalised with respect to the branching fraction of $\Bd\to\jpsi\Kstarz$. 

The systematic uncertainties on $A_{FB}$, $F_{L}$ and the differential branching fraction are typically 30\% of the statistical uncertainty. Across $q^{2}$ a dominant contribution to the systematic uncertainty comes from the understanding of data-derived corrections to the detector performance, which is expected to improve with increased statistics. The choice of the background angular and mass model in the fit for $A_{FB}$ and $F_{L}$ lead to a systematic uncertainty at the level of 10-20\% of the statistical uncertainty. This model dependence will also improve with an enlarged data set. 

The observed forward-backward asymmetry, longitudinal \Kstarz polarisation and differential branching fraction are consistent with a SM interpretation. In particular the observed forward-backward asymmetry shows evidence for a zero-crossing point which could be measured with a larger data set.

\section{Radiative decays in \lhcb}
\label{sec:btosgam}

The long term goal of the \lhcb radiative decay program is a measurement of the right-handed component of the photon polarization, probed using a time dependent analysis of $\Bs\to\phi\g$ decays \cite{bib:zwicky}. In the shorter term, a large \BdKsg yield will enable \lhcb to make a world leading measurement of the $\mathcal{A}_{CP}(\BdKsg)$.  The most precise single measurement of $\mathcal{A}_{CP}(\BdKsg)$ comes from the \babar experiment, based on 2400 \Bd candidates \cite{bib:babaracp}. A large $\Bs\to\phi\g$ could allow an improved measurement of the $\Bs\to\phi\g$  branching fraction.

\begin{figure}[!ht]
\centering
\includegraphics[width=80mm]{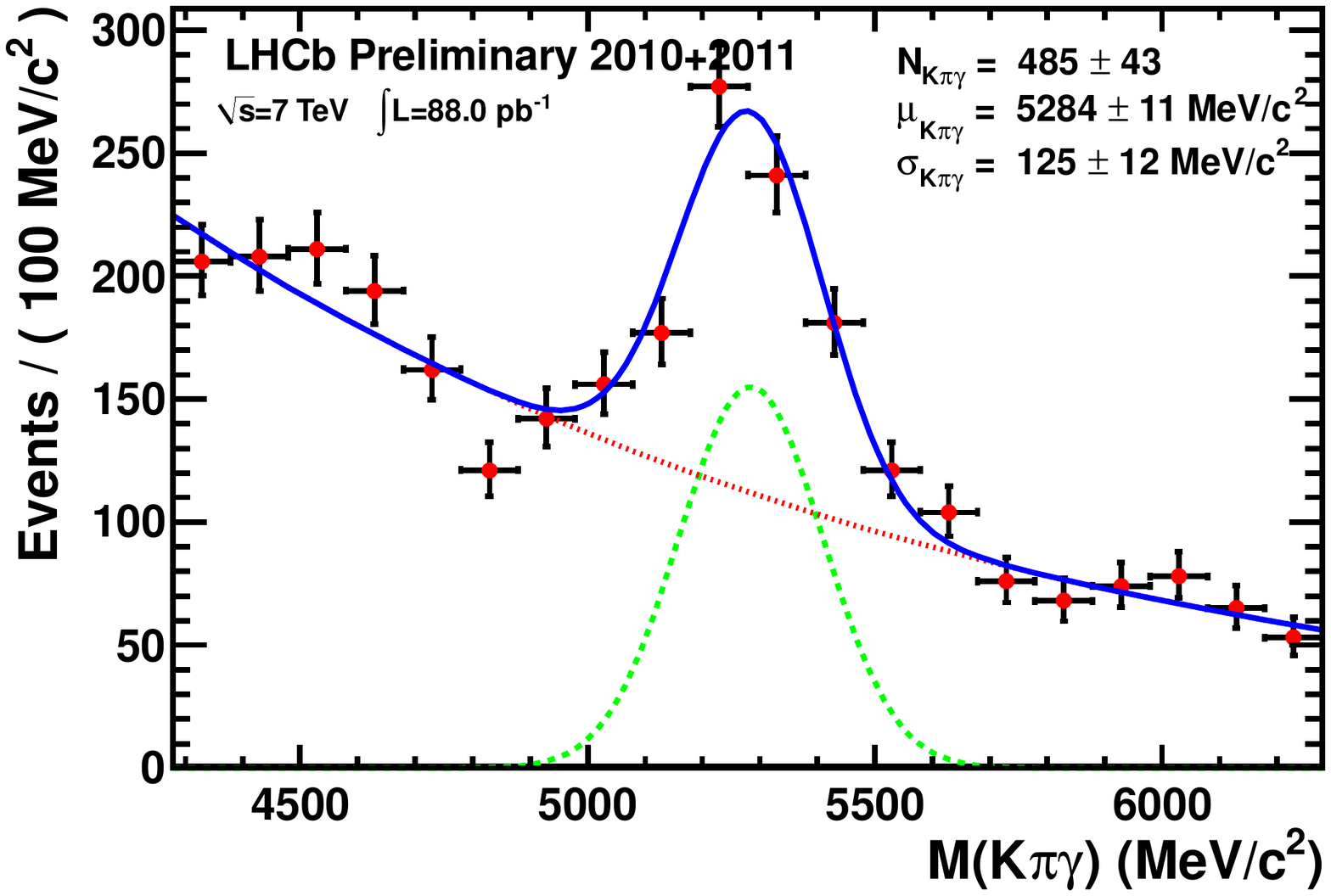} 
\includegraphics[width=80mm]{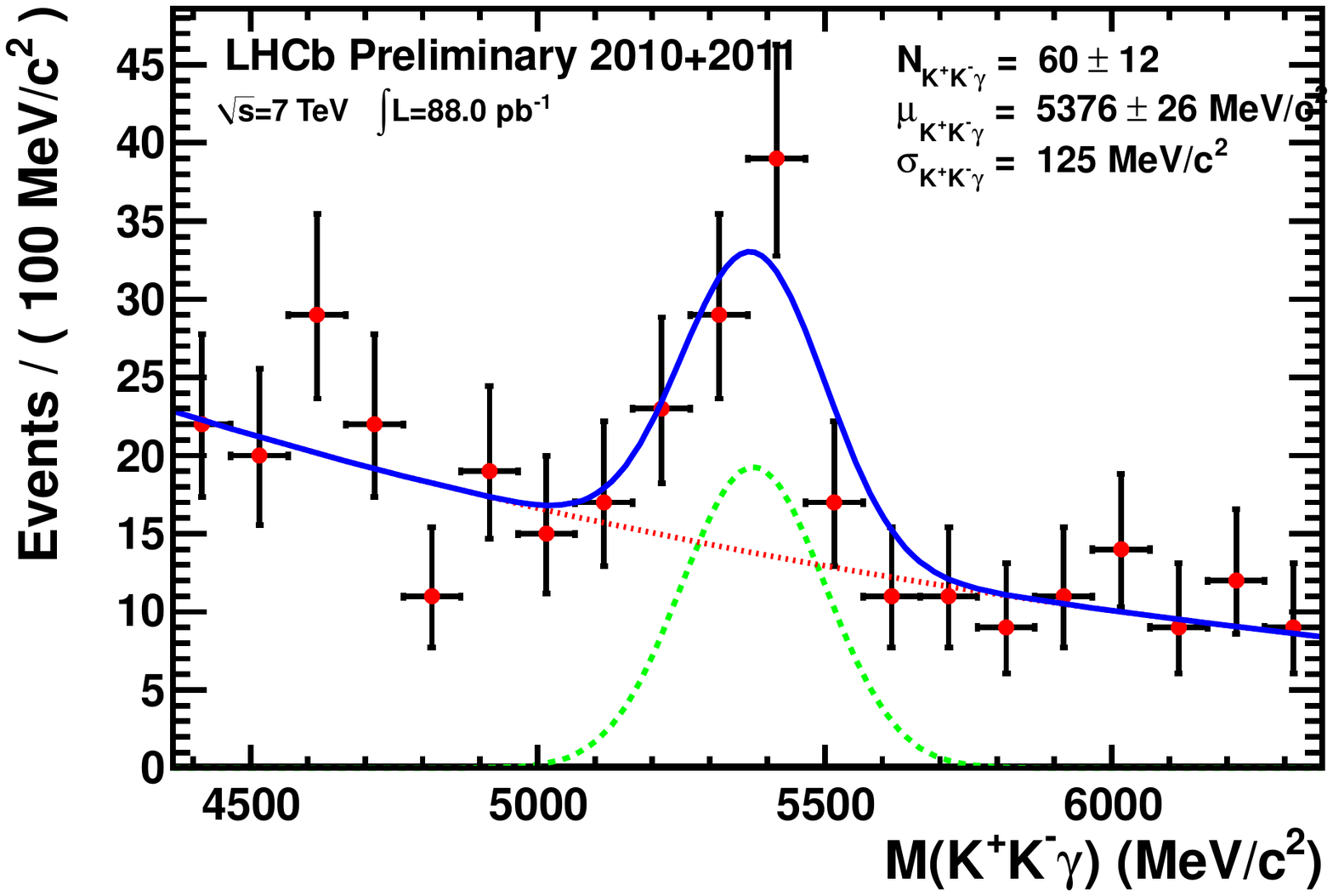} 
\caption{Reconstructed ${\Kpm\pimp\g}$ (top) and ${\Kpm\Kmp\g}$ mass distributions in 88\invpb of integrated luminosity. The solid line is a fit to the data with a Gaussian signal distribution (green-dotted line) and exponential background distribution (red-thin line) \label{fig:btosg:mass}}
\end{figure}

In 88\invpb collected by \lhcb in 2010 and 2011, \lhcb observes $485\pm43$ \BdKsg candidates and $60\pm12$ $\Bs\to\phi\g$ candidates. The $\Bs\to\phi\g$ yield is the largest yield of $\Bs$ radiative decays at a single experiment. The ${\Kpm\Kmp\g}$ and ${\Kpm\pimp\g}$ mass distributions of these candidates is shown in Fig.~\ref{fig:btosg:mass}. The reconstructed ${\Kpm\Kmp\gamma}$ and ${\Kpm\pimp\gamma}$ signal mass resolution is $\sim125\mevcc$. This is larger than the Monte Carlo prediction of 100\mevcc, but will improve with ongoing (time-dependent) Calorimeter calibration.

\BdKsg and $\Bs\to\phi\g$ candidates are triggered in \lhcb's Level-0 hardware trigger on the high-$E_{T}$ photon or by the kaon or pion from the $\phi$ or \Kstarz ~\cite{bib:level0}. In the first stage of the software trigger the events are triggered by a high-$p_{T}$ and high-IP track (from the $\phi$ or \Kstarz) or by a softer track and a high-$E_{T}$ photon. In the second stage of the software trigger a full exclusive reconstruction of the \B is performed which mirrors the offline selection. 

The offline selection is similar to the pre-selection of \BdKsmm described above in that the tracks from the \Kstarz and $\phi$ are required to have significant impact parameter and the \B is required to point back to the PV. Additional requirements are then made of the \g \pt ($\pt > 2.6\gevc$) and on the \B \pt ($\pt > 3\gevc$) and isolation to improve the purity of the signal. 

Extrapolating to a 1\invfb data set, \lhcb expects to collect a data sample of $\mathcal{O}(5000)$ \BdKsg candidates and $\mathcal{O}(700)$ $\Bs\to\phi\g$ candidates, which would also represent the largest sample of radiative \Bd decays at a single experiment. 

\section{Summary}

In 2011 the performance of the LHC and \lhcb has been excellent delivering 300\invpb of integrated luminosity in the first three months of data taking.  This data sample has enabled \lhcb to make the world's most precise measurement of the $A_{FB}(q^{2})$ in \BdKsmm, which shows a striking agreement with the SM prediction. By the end of 2011, it is expected that \lhcb will have recorded 1\invfb of integrated luminosity providing the worlds largest samples of reconstructed \BdKsg and $\Bs\to\phi\g$ decays. This expanded data set will also enable \lhcb to make a first measurement of the forward-backward asymmetry zero-crossing point and transverse asymmetries~\cite{bib:kstarmumu:full} in \BdKsmm and to pursue a wide range of other rare decay measurements.   


\begin{acknowledgments}

We express our gratitude to our colleagues in the CERN accelerator
departments for the excellent performance of the LHC. We thank the
technical and administrative staff at CERN and at the LHCb institutes,
and acknowledge support from the National Agencies: CAPES, CNPq,
FAPERJ and FINEP (Brazil); CERN; NSFC (China); CNRS/IN2P3 (France);
BMBF, DFG, HGF and MPG (Germany); SFI (Ireland); INFN (Italy); FOM and
NWO (Netherlands); SCSR (Poland); ANCS (Romania); MinES of Russia and
Rosatom (Russia); MICINN, XUNGAL and GENCAT (Spain); SNSF and SER
(Switzerland); NAS Ukraine (Ukraine); STFC (United Kingdom); NSF
(USA). We also acknowledge the support received from the ERC under FP7
and the R\'egion Auvergne.

\end{acknowledgments}

\bigskip 

\end{document}